\documentclass[osajnl,twocolumn,showpacs,superscriptaddress,10pt]{revtex4-1} %% use 11pt for Applied Optics
\usepackage{amsmath,amssymb,graphicx, subfig}
\begin{document}

\title{Characterising Conical Refraction Optical Tweezers }

\author{C. McDonald} \email{Corresponding author: c.y.mcdonald@dundee.ac.uk}
\author{C. McDougall}
\affiliation{SUPA, Divsion of Physics, University of Dundee, Nethergate, Dundee, DD1 4HN UK}
\author{E. Rafailov}
\affiliation{Institute of Photonic Technologies, Aston University, Birmingham, B4 7ET, United Kingdom, }
\author{D. McGloin}
\affiliation{SUPA, Divsion of Physics, University of Dundee, Nethergate, Dundee, DD1 4HN UK}

\begin{abstract}
Conical refraction occurs when a beam of light travels through an
appropriately cut biaxial crystal. By focussing the conically refracted
beam through a high numerical aperture microscope objective, conical
refraction optical tweezers can be created, allowing for particle
manipulation in both Raman spots and in the Lloyd/Poggendorff rings.
We present a thorough quantification of the trapping properties of
such a beam, focussing on the trap stiffness and how this varies with
trap power and trapped particle location. We show that the lower Raman
spot can be thought of as a single-beam optical gradient force trap,
while radiation pressure dominates in the upper Raman spot, leading
to optical levitation rather than trapping. Particles in the Lloyd/Poggendorff
rings experience a lower trap stiffness than particles in the lower
Raman spot but benefit from rotational control.
\end{abstract}

% REPLACE WITH CORRECT OCIS CODES FOR YOUR ARTICLE
% NOTE: \ocis{} IS ALIASED TO \pacs{} BUT MUST
% FORMAT THE TERMS CORRECTLY FOR EACH JOURNAL
\maketitle %% required
Traditionally, optical tweezers have been created by focussing a single
beam of light, with a Gaussian intensity distribution, through a high
numerical aperture (NA) lens ~\cite{Ashkin1986}. This allowed microscopic
particles to be trapped and moved, and has provided insight in to the physical,
biological and chemical processes occurring in single molecules \cite{Svoboda1993},
aerosols \cite{Burnham2006} and liquids \cite{Pralle1998}. More
recently, variations on single beam optical traps, based on beam shaping
techniques, have appeared, further exploiting light's ability to exert
a force on objects: Bessel beams were used
to guide particles over long distances \cite{Arlt2001}; rotational
control of particles was demonstrated \cite{Paterson2001}; and computer
generated holograms were used to create multi-beam traps \cite{Grier:06}.

Beam shaping techniques, of course, have a much longer history than
those developed for optical tweezers. One of the earliest
manifestations of significant beam shaping is that of conical refraction,
first proposed by William Hamilton in 1832 \cite{Hamilton1837}. Conical
refraction occurs upon passing light through an appropriately cut
biaxial crystal and produces a beam of light that propagates on a
dual-cone \cite{Rafailov2013}, resulting in the creation of a ring structure, called Lloyd/Poggendorff
rings, at a point beyond the crystal \cite{Berry2006}. On either
side of the Lloyd/Poggendorff rings are the Raman spots, which are
more akin to rods of light than spots \cite{Peet2013150}.

\begin{figure}
\includegraphics[width=1\columnwidth]{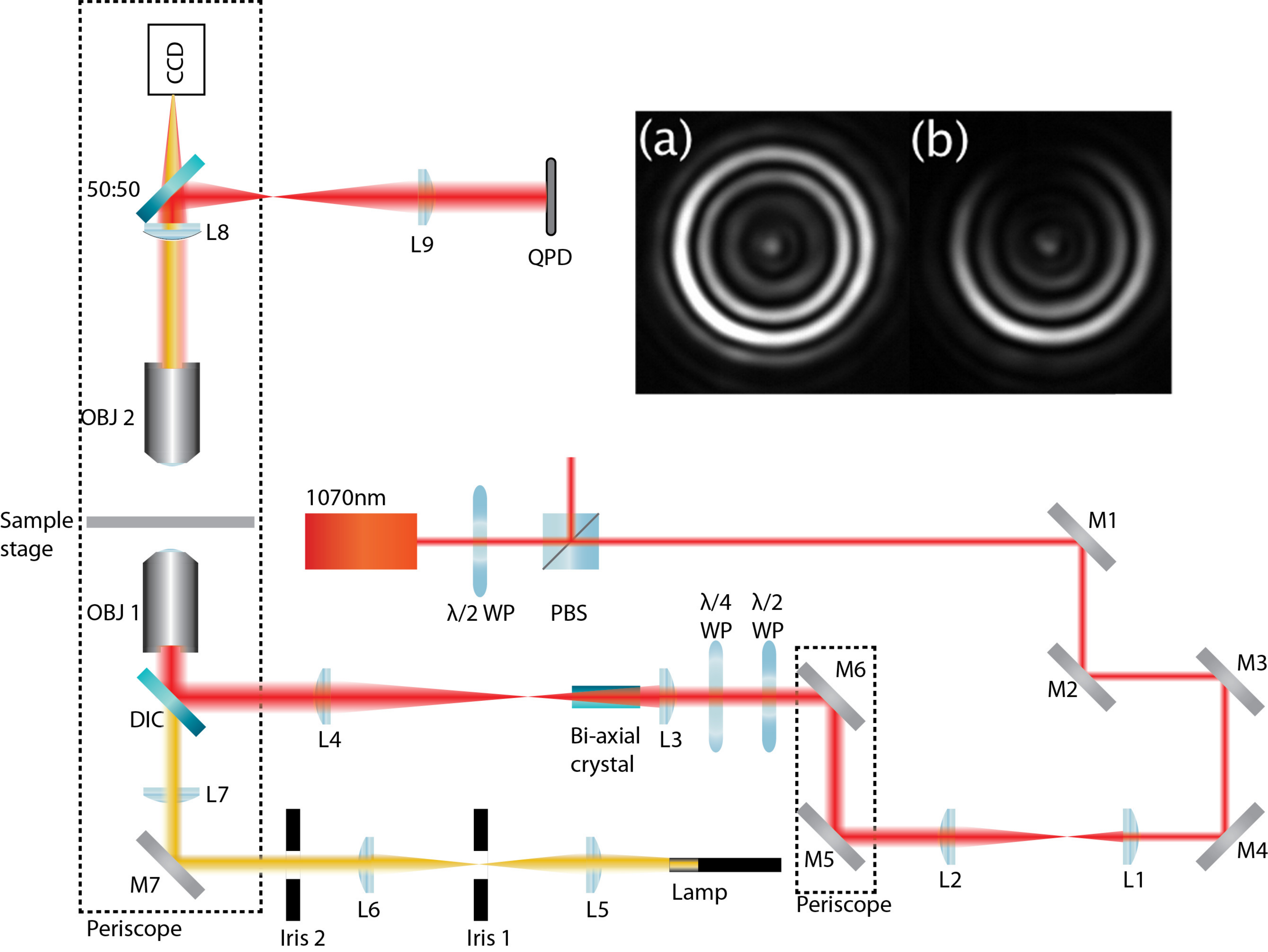}
\protect\caption{Experimental set-up for conical refraction tweezers. $\lambda/2$
WP = half wave plate, $\lambda/4$
WP = quarter wave plate, PBS = polarising beam splitting cube, DIC
= Dichroic mirror, OBJ 1 = Nikon 0.9NA 50x oil immersion objective,
OBJ 2 = Nikon 1.25NA 100x oil immersion objective, L\# = lens, M\# = mirror. Not shown: sample
stage was positioned on a Newport xyz translation stage; OBJ 2, L8,
L9, 50:50 splitter and CCD camera were positioned on a 2nd Newport
xyz translation stage; QPD was positioned on another Newport xyz translation
stage. Inset shows Lloyd/Poggendorff rings for (a) circular and (b)
linear incident beam polarisations. Out of focus light is visible within, and surrounding, the Lloyd/Poggendorff rings due to the depth of field of the imaging system. \label{fig:set-up}}
\end{figure}

Conically refracted beams have been used in optical manipulation studies \cite{ODwyer2010,CMcDougall2012,Turpin2013,ODwyer2012} but their trapping properties have not been fully investigated. Here, we use power spectral analysis \cite{Berg-Sorensen2004} to quantify the trapping ability of a conically refracted beam (where we assume the beam is approximated by a harmonic potential well) by calculating the trap stiffness \cite{Block2004} of the three main parts of the beam: the upper and lower Raman spots and the Lloyd/Poggendorff rings.
\\
\indent Our experimental set-up is shown in Fig.\ref{fig:set-up} and makes use of a 1W (maximum output) 1070nm fibre laser (Model: PYL-1-1064-LP, IPG Photonics) as the trapping source. The biaxial crystal is $KGd(WO_4)_2$ 12mm in length. The facet of the crystal is orientated perpendicular to the beam propagation axis, with the beam focused to a point beyond the crystal. The main optics are set up in a straight line (the optics focussing the laser through the crystal and the imaging optics up to the dichroic beamsplitter) in an attempt to avoid any unwanted polarisation effects in the optical train, which lead to a non-uniform ring pattern \cite{CMcDougall2012}. Removable half-wave and quarter-wave plates are included to allow the input polarisation to the crystal to be selectively changed from linear to circular polarisation. Transmission microscopy, where the sample is illuminated from below and imaged from above, is used as previous work \cite{CMcDougall2012} has shown differences in beam spot appearance when imaged in transmission and reflection mode. A Nikon 0.9NA 50x oil immersion objective is used to focus the conical beam and form the trap, while a Nikon 1.25NA 100x oil immersion is used to image the sample onto the CCD camera and to image the beam onto a quadrant photodiode (QPD). The 10mm diameter, silicon QPD is used in back focal plane interferometry (BFPI) mode \cite{Block2004} and is connected to custom built transimpedance amplifiers, which are, in turn, connected to a National Instruments SCB-68A connector block. Signals are collected via a National Instruments PCI-6250 DAQ card and analysed using an in-house LabVIEW program. By mounting the imaging system and the QPD on separate xyz translation stages, the focussed beam can be axially scanned, allowing for the full beam profile to be imaged by the camera, and the QPD can be realigned to the new position.

For the trap stiffness measurements, the camera was first aligned
to where the trapped particle would be in focus and the QPD aligned
so that all four quadrants produced the same voltage. The particle
was then trapped and the QPD sampled, in four second windows, at a
rate of 50kHz. Power spectra were calculated from the average signal
of five of these ``windows'' and the trap stiffness in X and Y ($k_x$ and $k_y$) was determined from
the average corner frequency of ten power spectra \cite{Berg-Sorensen2004}.

% \section{Results and Discussion}

\begin{figure}
%\subfloat[]{\includegraphics[width=1\columnwidth]{Figure2a.eps}}
%
%\subfloat[]{\includegraphics[width=1\columnwidth]{Figure2b.eps}}
%
%\subfloat[]{\includegraphics[width=0.33\columnwidth]{Figure2c.eps}}
%
%\subfloat[]{\includegraphics[width=0.33\columnwidth]{Figure2d.eps}}
%
%\subfloat[]{\includegraphics[width=0.33\columnwidth]{Figure2e.eps}}

%\subfloat[]{\includegraphics[width=1\columnwidth]{Figure2a2-eps-converted-to.pdf}
%
%}
%
%\subfloat[]{\includegraphics[width=1\columnwidth]{Figure2b2-eps-converted-to.pdf}
%
%}
%
%\subfloat[]{\includegraphics[width=0.33\columnwidth]{Figure2c2-eps-converted-to.pdf}
%
%
%
%}\subfloat[]{\includegraphics[width=0.33\columnwidth]{Figure2d2-eps-converted-to.pdf}
%
%
%
%}\subfloat[]{\includegraphics[width=0.33\columnwidth]{Figure2e2-eps-converted-to.pdf}
%
%
%
%}
\includegraphics[width=1\columnwidth]{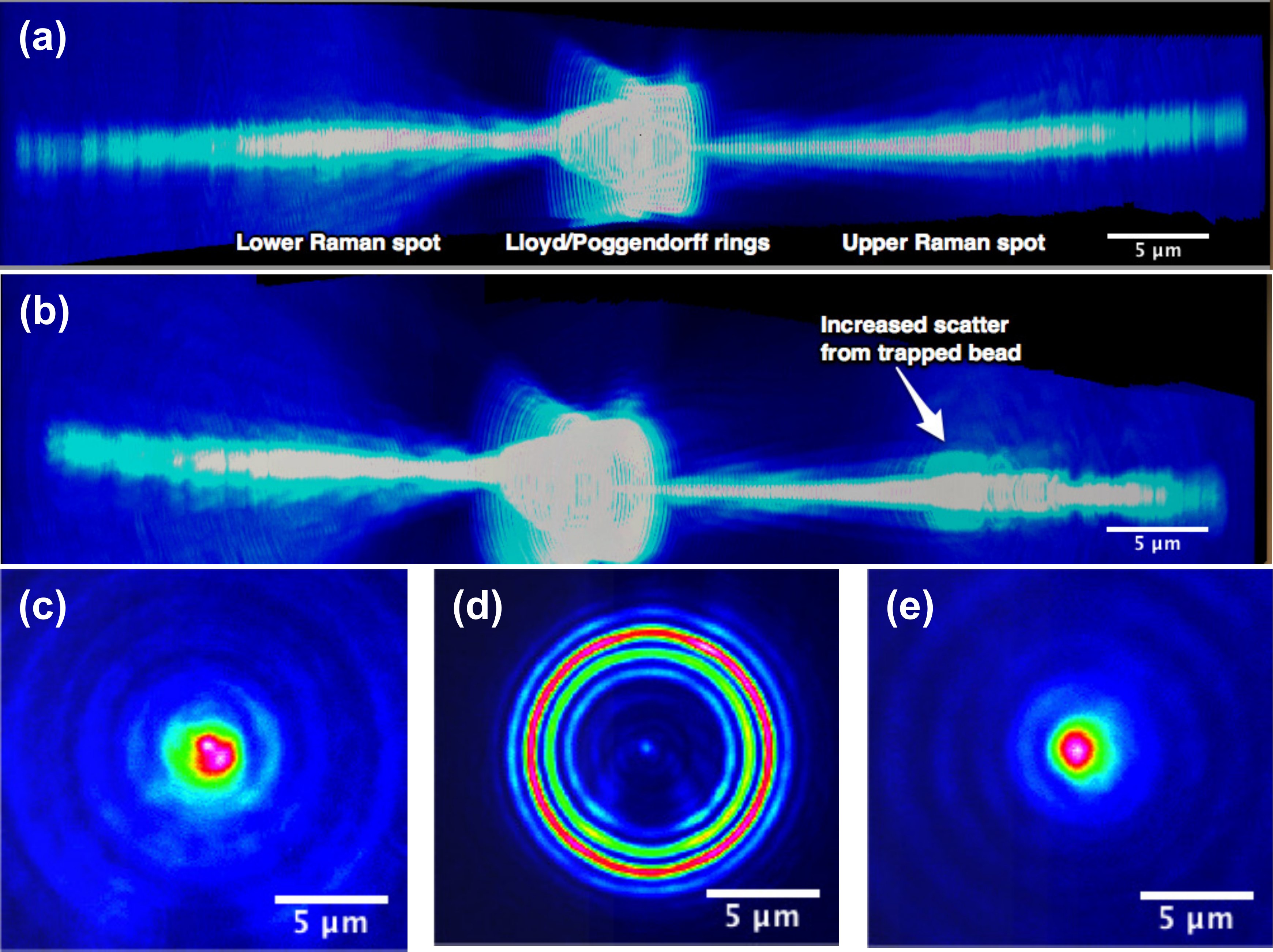}
\protect\caption{Circularly polarised, conically refracted beam, \~75.5$\mu$m
long, propagating from left to right, focussed with a
0.9NA objective. (a) Empty trap. (b) 2.56$\mu$m
bead in upper Raman spot, identifiable by increased scatter. The slight asymmetry in the images is due to the projection method used. Cross
sections of the lower Raman spot (c), Lloyd/Poggendorff rings (d) and upper
Raman spot (e) are shown for
an empty trap.\label{fig:Beam-Profile}}
\end{figure}

Initially, the shape of the trapping beam was investigated to ensure
a fully conically refracted beam was generated in the trapping plane.
A beam profiler was placed in a conjugate plane to the CCD camera
and the translation stage was used to scan the beam axially. By saving
sequential images of the beam from the beam profiler as the imaging
system is moved in 0.25$\mu$m steps, a full 3D scan of the conically refracted beam can be generated.
Fig.\ref{fig:Beam-Profile} shows an example of the z-cross section
of the beam under high NA focussing for circular input polarisation,
with and without a trapped bead. If the beam incident on the crystal
is linearly polarised, a segment from the Lloyd/Poggendorff rings
will be missing and the intensity will change along the ring, Fig.\ref{fig:set-up} inset. This missing region, located parallel to the input polarisation, would have had polarisation
perpendicular to that of the incident beam, had the beam contained
the corresponding electric field component \cite{Berry200713}. Trapped particles, in this case, would move to the region of highest intensity, opposite the missing segment of the beam.

\begin{figure*}
%\subfloat[]{\includegraphics[width=0.45\columnwidth]{Figure3a.eps}}
%
%\subfloat[]{\includegraphics[width=0.45\columnwidth]{Figure3b.eps}}
%
%\subfloat[]{\includegraphics[width=0.45\columnwidth]{Figure3c.eps}}
%
%\subfloat[]{\includegraphics[width=0.45\columnwidth]{Figure3d.eps}}
%
%\subfloat[]{\includegraphics[width=0.45\columnwidth]{Figure3e.eps}}
%
%\subfloat[]{\includegraphics[width=0.45\columnwidth]{Figure3f.eps}}
%

%\subfloat[]{\includegraphics[width=0.5\columnwidth]{Figure3a4.pdf}
%}
%\subfloat[]{\includegraphics[width=0.5\columnwidth]{Figure3b4.pdf}
%}
%
%\subfloat[]{\includegraphics[width=0.5\columnwidth]{Figure3c4.pdf}
%}
%\subfloat[]{\includegraphics[width=0.5\columnwidth]{Figure3d4.pdf}
%}
%
%\subfloat[]{\includegraphics[width=0.5\columnwidth]{Figure3e4.pdf}
%}
%\subfloat[]{\includegraphics[width=0.5\columnwidth]{Figure3f3.pdf}
%}
\includegraphics[width=1\textwidth]{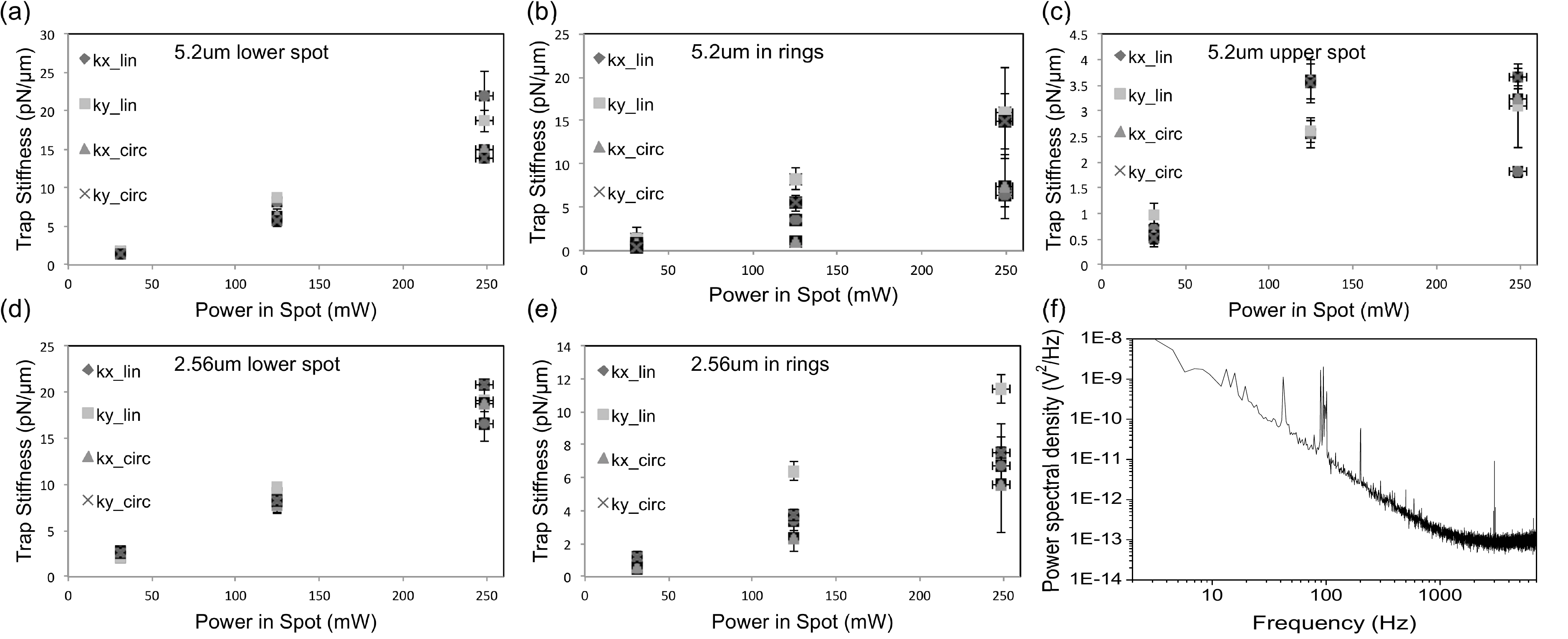}
\protect\caption{Trap stiffness vs power for:
(a) 5.2$\mu$m bead
and (b) 2.56$\mu$m
bead trapped in lower Raman spot; (c) 5.2$\mu$m
bead and (d) 2.56$\mu$m
bead trapped in the Lloyd/Poggendorff rings; (e) 5.2$\mu$m
bead trapped in upper Raman spot. kx lin, ky lin, kx circ and ky circ denote, respectively, trap stiffness in X and Y for linear and circular polarised light. (f) Power spectrum for a 2.56$\mu$m
bead trapped in upper Raman spot.\label{fig:Trap-stiffness-graphs}}
\end{figure*}

Linear and circular input polarisations were used to trap both 2.56$\mu$m
and 5.2$\mu$m silica
beads, with various trapping powers and in different locations of
the beam. 2.56$\mu$m beads were chosen as they are comparable in size to the lower Raman spot, with 5.2$\mu$m beads roughly comparable to the radius of the Lloyd/Poggendorff rings. The trap stiffness for each of these situations was investigated
and is shown in Fig.\ref{fig:Trap-stiffness-graphs}.

% \subsection{Trap Stiffness as a Function of Bead Location}

Trapping in the lower Raman spot shows the highest trap stiffness
for the beam, for both polarisations and for both
particle sizes, Figs.\ref{fig:Trap-stiffness-graphs}(a) and \ref{fig:Trap-stiffness-graphs}(b).
The small error bars in these graphs also suggest that, for both sizes
and polarisation states, a stable, three dimensional, trap is formed.
Both beads, in this case, are trapped at the top of the lower Raman
spot, before it evolves into the ring planes. Fig.\ref{fig:Beam-Profile}(a)
shows a slight focussing of the beam at this point, indicating that
there could be a peak in the gradient force here and, therefore, increased
trapping performance. A 5.2$\mu$m bead trapped with 125.1$\pm$ 1.9 mW at
the top of the lower Raman spot, with linear polarisation, has a trap
stiffness of 8.2$\pm$0.7pN/$\mu$m, Fig.\ref{fig:Trap-stiffness-graphs}(a).
This is significantly lower than the trap stiffness of 49.4$\pm$12.6pN/$\mu$m
that is achievable for the same particle under similar conditions
but trapped with a Gaussian shaped trap, discussed later in the paper.

% shown in subsection \ref{sub:Levitation-Experiments}.

Trapping of both bead sizes can be achieved in the Lloyd/Poggendorff
rings but with a lower trap stiffness than that of trapping in the
lower Raman spot, Figs.\ref{fig:Trap-stiffness-graphs}(c) and
\ref{fig:Trap-stiffness-graphs}(d). However, although there is a
decrease in trapping performance, this is outweighed by the additional,
rotational control which is gained. By rotating the waveplates, it
is possible to rotate the trapped particles around the circumference
of the rings\cite{ODwyer2010, CMcDougall2012}.

Trapping in the upper Raman spot shows a very small trap stiffness
for the 5.2$\mu$m particles,
indicating levitation rather than trapping, Fig.\ref{fig:Trap-stiffness-graphs}(e).
However, power spectra for 2.56$\mu$m
beads in the upper Raman spot, Fig.\ref{fig:Trap-stiffness-graphs}(f),
were indicative of an empty trap. The larger particles appear more
stable in the upper Raman spot, while small particles may sink and
rise with laser power fluctuations. The power spectrum shown in Fig.\ref{fig:Trap-stiffness-graphs}(f) falls off as $D/\pi^2 f^2$, where $D$ is the diffusion coefficient for a particle in water\cite{Berg-Sorensen2004}.
This is characteristic of free diffusion, therefore the particle is
moving as if the trap did not exist. This indicates that the particle
is experiencing an approximately flat trapping potential and is free
to diffuse over a volume that is comparatively large to the particle
diameter. Due to the converging and diverging nature of the focussing envelope of the beam, at the top of the upper Raman spot we have a diverging beam, so we infer a lower potential gradient.

% \subsection{Trap Stiffness as a Function of Power}

Trap stiffness was investigated as a function of power, to further
quantify the trapping ability of a conically refracted beam. Figs.\ref{fig:Trap-stiffness-graphs}(a) and \ref{fig:Trap-stiffness-graphs}(b)
show that, as expected, trap stiffness increases linearly for increased
power. This is true for both input polarisation states and for both
particle sizes, indicating that a true 3D gradient trap is formed
at the top of the lower Raman spot.

Increasing the power also gives an increase in trap stiffness for
particles trapped in the Lloyd/Poggendorff rings. However, it appears
that there is an asymmetry in the X and Y trap stiffnesses for linear
input polarisation, which becomes more pronounced with increasing
trap power, Figs.\ref{fig:Trap-stiffness-graphs}(c) and \ref{fig:Trap-stiffness-graphs}(d). This asymmetry in the trap stiffnesses could be attributed to this ``missing''
polarisation contribution from the incident beam.

The larger, 5.2$\mu$m
beads, experience very little change in their trap stiffness with
increasing power when trapped in the upper Raman spot, Fig.\ref{fig:Trap-stiffness-graphs}(e).
The 2.56$\mu$m beads
did not show any change in their power spectra with increasing power.
Even at the highest trapping power used, 248.9$\pm$5.2mW, when a traditional, Gaussian trap would
have its highest trap stiffness, the power spectra for a trapped 2.56$\mu$m
bead had the characteristic shape of free diffusion of the particle \cite{Berg-Sorensen2004},
further indication that levitation is occurring.

% \subsection{Levitation Experiments\label{sub:Levitation-Experiments}}

\begin{figure}
\includegraphics[width=0.75\columnwidth]{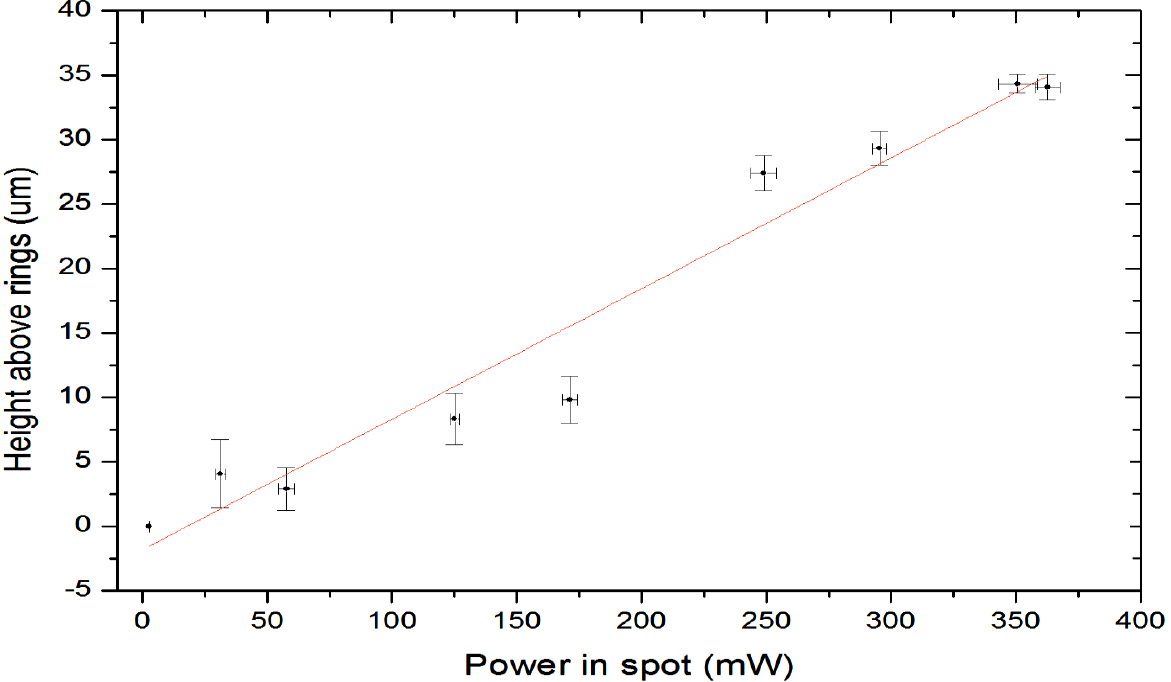}
\protect\caption{5.2$\mu$m bead in upper Raman spot, height above Lloyd/Poggendorff
ring plane vs trap power.\label{fig:5.2=00003D00003D0003BCm-bead-height}}
\end{figure}

In order to confirm that levitation was occurring in the upper Raman
spot, bead position was investigated as a function of power. Beads
were trapped in the upper Raman spot and brought in to focus on the
CCD camera. The trap power was varied between 2.6$\pm$0.1mW and
362.7$\pm$4.9mW, with the imaging system moved to bring the bead
back in to focus at each new power. Fig.\ref{fig:5.2=00003D00003D0003BCm-bead-height}
shows how the height above the Lloyd/Poggendorff rings for a 5.2$\mu$m
bead in the upper Raman spot increases with increasing trap power.
This is characteristic of optical levitation and, combined with the
weak trap stiffnesses shown in Fig.\ref{fig:Trap-stiffness-graphs}(e)
and the power spectrum shown in Fig.\ref{fig:Trap-stiffness-graphs}(f), 
can be taken as proof that optical levitation, as opposed to optical trapping, is occurring in
the upper Raman spot. The same experiment was performed for beads
trapped in the lower Raman spot and in the Lloyd/Poggendorff rings
but there was no observable change in axial position of the bead with
changing trap power, further indicating that a 3D gradient force must
be present in these locations.

\begin{figure}
\includegraphics[width=0.75\columnwidth]{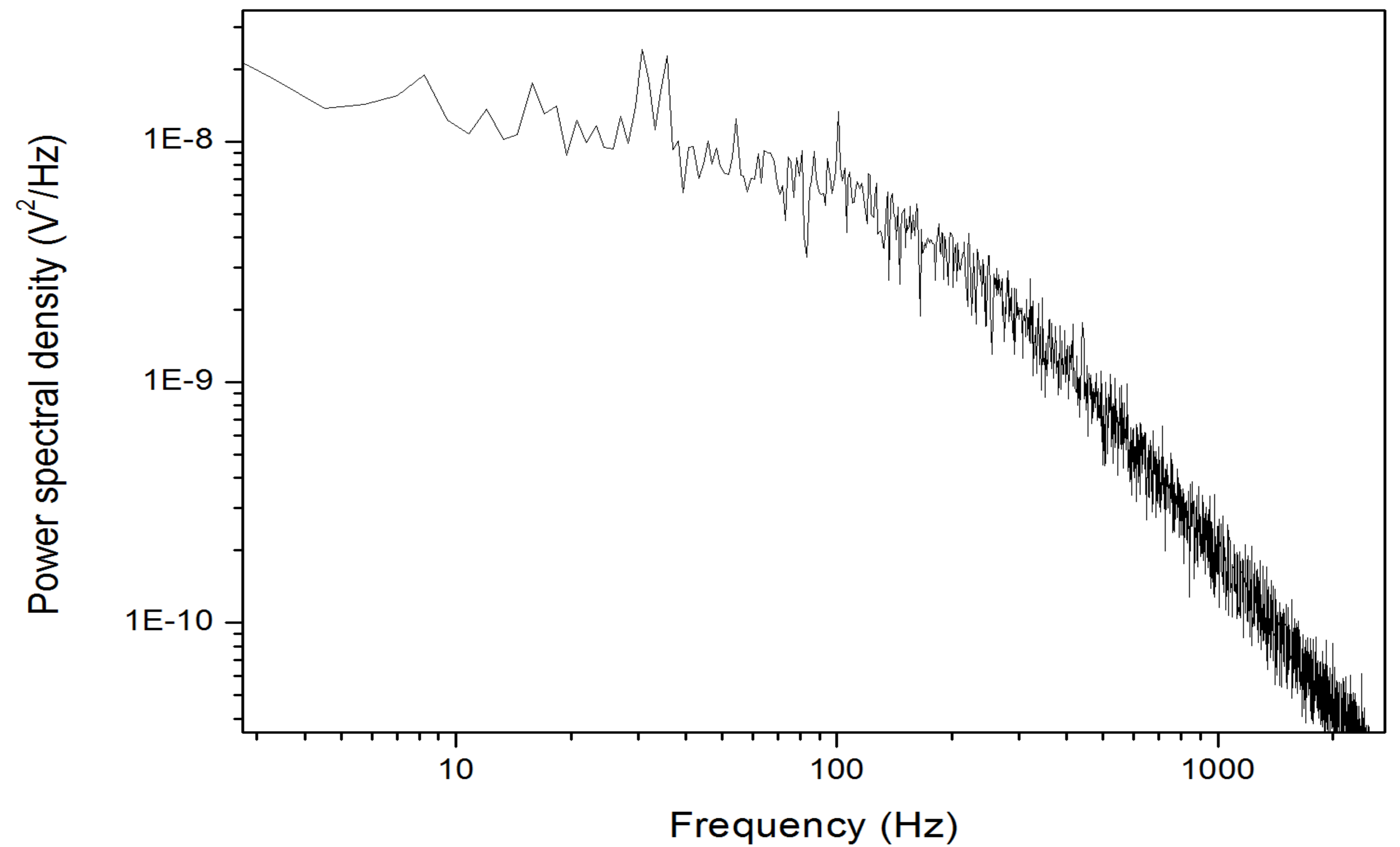}\protect\caption{Power spectrum for a 5.2$\mu$m
bead trapped with 137$\pm$6mW in a Gaussian trap.\label{fig:Power-spectrum}}
\end{figure}

To ensure that the levitation is not due to the comparatively low
numerical aperture of the trapping objective, the Nikon 0.9NA 50x
oil immersion objective was used on a traditional set of optical tweezers.
Trapped bead position was shown to not change as a function of power,
indicating that three dimensional gradient trapping was occurring.
A 5.2$\mu$m bead was
trapped with a Gaussian beam and a trap power of 137$\pm$6mW and
power spectra calculated \cite{Berg-Sorensen2004}, an example 
is shown in Fig.\ref{fig:Power-spectrum}. An average trap stiffness
of 49.4$\pm$12.6pN/$\mu$m
was determined, which, combined with particle position being independent
of trap power, shows that gradient trapping is possible with a 0.9NA
objective and that, therefore, the optical levitation observed is
due to the conically refracted beam and not the low numerical aperture
objective. A 1.25NA oil objective was used to generate the conical trap, to try and extend the trapping limit of the upper Raman spot. However, this produced a pattern which was too compact to trap, separately, in all three components of the beam.

% \section{Conclusion}

By quantifying the trapping properties of a conically refracted beam,
through power spectrum analysis \cite{Berg-Sorensen2004}, the three trapping regimes
of the beam have been investigated. The lower Raman spot has been
shown to have good trapping ability, with a comparatively high trap
stiffness, and so can function as a traditional, gradient force, optical
tweezers. Rotational control of the particle can be achieved by trapping
in the Lloyd/Poggendorff rings, where the trap stiffness is still
sufficient to trap the particle in all three dimensions. Finally,
the particle can be guided along the upper Raman spot by varying the
trap power, thanks to its levitation properties. Recent work concerning
the conical diffraction of azimuthally and radially polarised light
\cite{Grant2014} has highlighted the importance of polarisation in
conical diffraction and its impact on beam shape. The trapping properties
of these beams remain unexplored and could warrant further investigation.
Applications where trapping, rotation and guiding are required could,
therefore, be achieved by using a conically refracted beam, negating
the need for complex techniques using, for example, spatial light
modulators. The study of photophoretic manipulation of light-absorbing
particles could benefit from such conical refracted beams, with confinement
possible in the dark regions of the beam enclosed by the Lloyd/Poggendorff rings \cite{Esseling2012}.    The hollow cone at the centre of the beam could find applications in the field of 3D STED microscopy \cite{Hell1994}.
% \begin{acknowledgments}

We thank Robert Henderson, Caren Kresse, Yu Loiko and David Carnegie for experimental
help. We thank EPSRC for support through grant EP/H004238/1 and for
the award of a studentship.
% \end{acknowledgments}
%\bibliographystyle{unsrt}
%\bibliography{ConicalRef}

\end{document}